# IMPROVING THE SCALABILITY BY CONTACT INFORMATION COMPRESSION IN ROUTING

V.Karthikeyan[1] and V.J.Vijayalakshmi[2]

[1]Department of ECE, SVS College of Engineering, Coimbatore, India
[2]Department of EEE, Sri Krishna College of Engg & Tech., Coimbatore, India

## ABSTRACT

*The existence of reduced scalability and delivery leads to the development of scalable routing by contact information compression. The previous work dealt with the result of consistent analysis in the performance of DTN hierarchical routing (DHR). It increases as the source to destination distance increases with decreases in the routing performance. This paper focuses on improving the scalability and delivery through contact information compression algorithm and also addresses the problem of power awareness routing to increase the lifetime of the overall network. Thus implementing the contact information compression (CIC) algorithm the estimated shortest path (ESP) is detected dynamically. The scalability and release are more improved during multipath multicasting, which delivers the information to a collection of target concurrently in a single transmission from the source.*

## KEYWORDS

*Delay Tolerant Network (DTN), DTN Hierarchical Routing (DHR), Contact Information Compression (CIC), Scalability and Delivery*

## 1. INTRODUCTION

The DTN Research Group (DTNRG) has designed architecture to support different protocols in DTN which can be described using graphs. So, any border in the graph indicates the continuation of set of contacts. The communication among any two nodes for a stage of instance is defined as the contact. Unlike types of contacts survived among which unrelenting contacts for incessant communication and predicted contacts for irregular communication amid nodes. listed contacts can stay alive, for occurrence, between a base station on earth and an Orbiting spread satellite. Opportunistic contacts are shaped just by the occurrence of two entities at the similar position. In the Opportunistic contacts a node waits to get together the goal in regulate to move its package. Messages transferred in DTNs are simply called bundles. They are transferred in an tiny manner between nodes using a transfer procedure that ensures node-to-node consistency. These messages can be of any size. Nodes are unspecified to contain huge buffers in which they can store up the bundles. The main benefit of this method is that it involves only one transmission per bundle. Most of the work relating to routing in DTN will be performed with predicted contacts. This DTN faces harsh problems due to the unreliable time and uncertainty in network connections. Hierarchical techniques are projected to improve scalability. In order to fully develop scalable routing techniques, it is essential to classify user communication sessions in a hierarchical way, so that the existing traffic prototype also scales well. The study of hierarchical routing reveals that, by adjusting the numeral of levels in the routing hierarchy a enhanced scalability can be achieved [2]. The DTN hierarchical routing address competent information aggregation and compression in the time-space domain while maintaining serious information Hierarchical routing requires the address of the source and the destination for efficient scalable routing [1]. If any of the sources in the network lacks in the address information, the source can remedy to a overhaul location.

## 2. RELATED WORKS

Routing in communication networks involves the indirection as of a unrelenting name (or ID) to a locator and delivering packets based in the lead the locator. On a large-scale, highly active network, the ID-to-locator mappings are both huge in number, and modify frequently. conventional routing protocols have need of elevated transparency to keep these indirections up-to-date [1]. In Weak State Routing (WSR), a routing device for large-scale highly active networks. WSR's innovation is that it uses arbitrary directional walks incomplete irregularly by weak indirection state information in midway nodes. The indirection position information is weak, i.e. interpreted not as complete truth, but as probabilistic hints. Nodes only have part information about the district a destination node is likely to be. This technique allows us to cumulative information about a number of distant locations in a geographic section. In other words, the state information maps a set-of-IDs to a geographical region [1]. In order to stay away from the propagation of the packets the threshold between 0.6 to 0.8 probabilities is permanent for each node and manage flooding can be avoided [4].

## 3. FORMULATED CIC ROUTING

In this proposed paper, we widen the study of scalable deterministic routing in DTNs by means of recurring mobility based on our preceding works. As an alternative of routing with global contact information, we proposed a routing algorithm that routes on contact information compacted by three collective methods. We address the challenge of well-organized information aggregation and firmness in the time-space area while maintaining significant information for capable routing.

## 4. DHR ALGORITHMS

Our proposed DHR is fairly clear-cut subsequent to the hierarchical network has been built. DHR is as well a hop by- hop routing. Every node makes its forwarding conclusion in two phases. The first phase runs only when the maximum level on which the cluster of the current node and that of the destination are different [1].

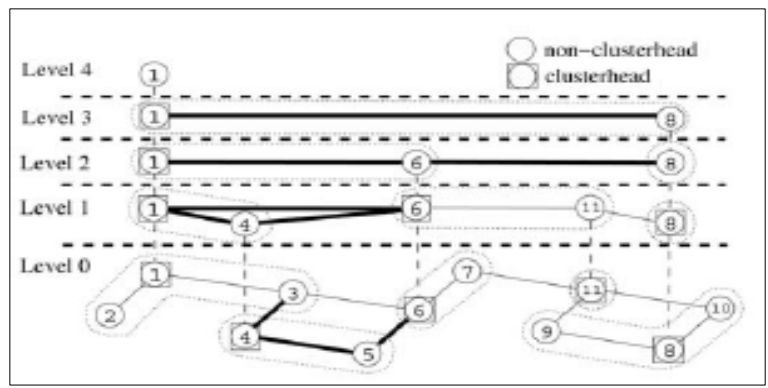

Fig.1

## 5.    MULTI PATH CREATION

It creates the nodes (sensors) according to the network capability. Here it will show in a draw panel. While the conception of, all the nodes show their connected calls and primarily it resolve be zero. After the construction of the multi paths, link between the nodes will set up and links are based on two circumstances, by finding adjacent neighbors and by linking to the inaccessible paths. Each path must join to the nodes through by the devices, according to the node competence and power efficient. If the network capability is less, there will be unsuccessful paths. User can see the listing of total nodes made.

## 6.    MULTI - PATH MULTICASTING

The proposed system is multicast data put on the air in multiple paths over wireless networks. We presume that the set of connections is evenly loaded, i.e., mobility and deprived channel circumstance rather than jamming are main reasons for packet crash. Begin by viewing the viability of numerous path multicasts, and subsequently move on to describe the way to forward the packets through multiple paths. The projected system has three basic steps, discovery of the shortest route, maintenance of the Route and Data Transmission [6].

### 6.1 Route Discovery

The underneath planning consists of 30 nodes in which two being source and destination others will be used for data transmission. The collection of a path for data broadcast is done based on the accessibility of the nodes in the section using the ad-hoc on insist remoteness vector routing algorithm. By using the Ad hoc on Demand Distance Vector routing protocol, the routes are created on demand, i.e. only when a route is needed for which there is no "fresh" record in the routing table [6].

### 6.2 Route Maintenance

The next step is the preservation of these routes which is evenly significant. The source has to incessantly observe the location of the nodes to create sure the data are creature agreed during the path to the destination lacking loss. In any case, if the location of the nodes change and the source doesn't make a note of it then the packets will be mislaid and ultimately have to be dislike [5].

### 6.3 Data Transmission

The pathway selection, preservation and data transmission are successive processes which happens in split seconds in real-time broadcast. Hence the paths allocated priory is used for data transmission. The first path allocated before is currently used for data transmission. The data is transferred through the tinted path. The second path chosen is now used for data transmission. The data is transferred through the highlighted path. The third path selected is used for data transmission. The data is transferred through the highlighted path [5].

## 7   MULTI-PATH POWER EFFICIENT ROUTING

A MANET might consist of nodes which are not capable to be re-charged in an predictable time period, energy maintenance is crucial to maintaining the lifetime of such a node. In networks consisting of these nodes, where it is unfeasible to stock up the nodes power, techniques for energy-efficient routing as well as competent data distribution between nodes is critical. An energy-efficient mechanism for input routing in sensor networks are called directed diffusion has been projected. Directed diffusion is an on-demand routing approach which has a data to drive occasionally and broadcasts it. When nodes receives a data, they send a support message to a pre-selected neighbor which indicates that it needs to receive more data from the selected neighbor. As these strengthening messages are propagated back to the source, an implied data path is set up; each midway node sets up state that forwards comparable data towards the previous hop [7].

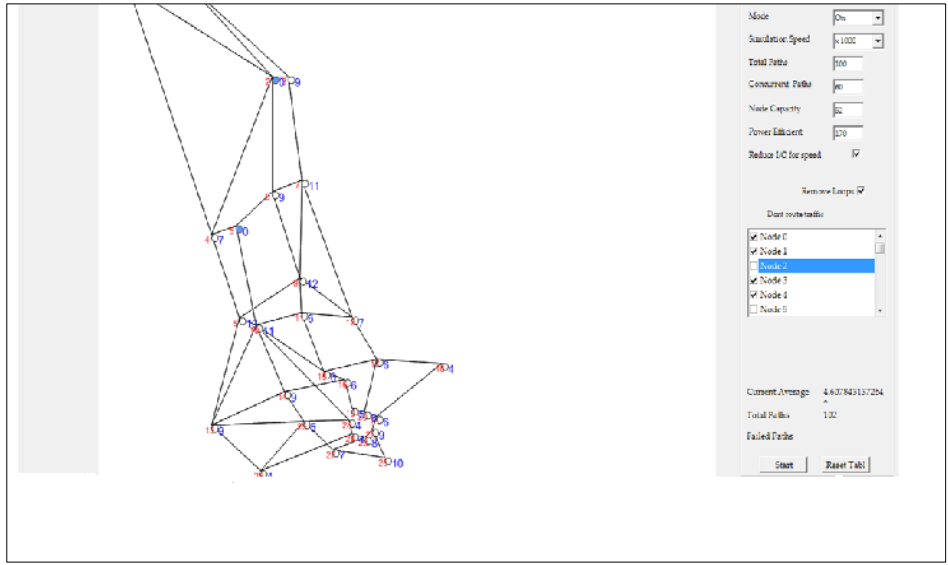

Fig.2

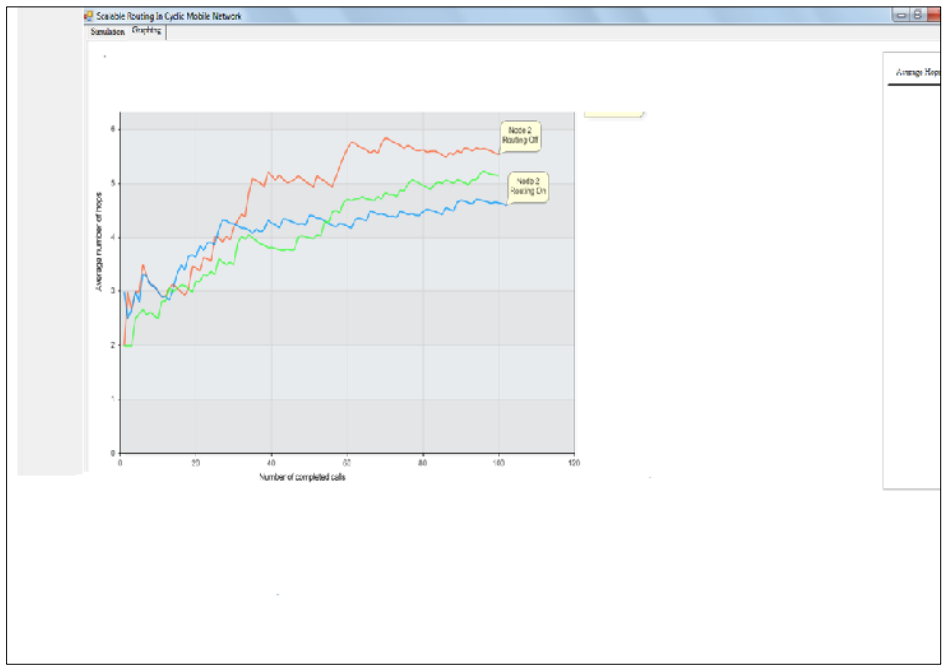

Fig.3

# 8   CONCLUSION

The projected work  has show in simulations results, the period of connection state measurements is chosen as one second. As a result, source nodes can keep informed their rates at most excellent roughly every two seconds, while it requires two measurements for estimating the slope vector according to the customized power algorithm. For simplicity the rate of redundancy is due to source coding, The most favorable values put forward that the difficulty of having elegant routers that are capable to ahead packets onto every branch at a dissimilar rate.  It offers only a insignificant profit in this situation, However, it is firm to depict any additional conclusions, as this result might depend on the exact topology and source-destination pair selections [4]. Also, our algorithm does better than traditional power algorithm as a effect of the accessibility of multiple trees to allocate the traffic load. However, while under network topology model the algorithm is capable to reduce the cost to a certain level, it cannot remove the packet losses and has a a great deal advanced overall cost compared to usual ones. The motive behind this result is the be short of of multicast functionality. Since we cannot generate multicast trees, the only savings due to multicasting occurs between the sources and overlay nodes.

**Authors**

**Prof. V. Karthikeyan** has received his Bachelor's Degree in Electronics and Communication Engineering from PGP college of Engineering and Technology in 2003, Namakkal, India, He received a Masters Degree in Applied Electronics from KSR college of Technology, Erode in 2006 He is currently working as Assistant Professor in SVS College of Engineering and Technology, Coimbatore. She has about 8 years of Teaching Experience

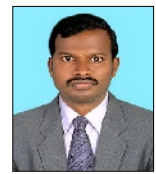

**Prof. V. J.Vijayalakshmi** has completed her Bachelor's Degree N Electrical & Electronics Engineering from Sri Ramakrishna Engineering College, Coimbatore, India. She finished her Masters Degree in Power Systems Engineering from Anna University of Technology, Coimbatore, She is currently working as Assistant Professor in Sri Krishna College of Engineering and Technology, Coimbatore She has about 5 years of Teaching Experience.

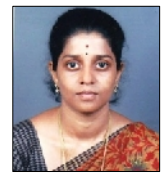